\documentclass[aps,prd,12point,twocolumn,nofootinbib,showpacs,superscriptaddress]{revtex4-1}
\def\theequation{\arabic{section}.\arabic{equation}}
\usepackage{psfrag}
\usepackage{subfigure}
\usepackage{color}
\usepackage{mathrsfs}
\usepackage{graphicx}
\usepackage{amssymb}
\usepackage{amsmath, amsthm}
\usepackage{epstopdf}
\usepackage{hyperref}
\usepackage{enumerate}

\newcommand{\be}{\begin{equation}}
\newcommand{\ee}{\end{equation}}

\begin{document}
\def\theequation{\arabic{section}.\arabic{equation}} 
%\preprint{}

\title{Simultaneous baldness and cosmic baldness and Kottler spacetime}

\author{Valerio Faraoni}
\email[]{vfaraoni@ubishops.ca}
%\thanks{}
\affiliation{Department of Physics and Astronomy, Bishop's 
University, 
2600 College Street, Sherbrooke, Qu\'ebec, Canada J1M~1Z7
}

\author{Adriana M. Cardini}
\email[]{acardini15@ubishops.ca}
%\thanks{}
\affiliation{Department of Physics and Astronomy, Bishop's 
University, 
2600 College Street, Sherbrooke, Qu\'ebec, Canada J1M~1Z7
}

\author{Wen-Jian Chung}
\email[]{wchung13@ubishops.ca}
%\thanks{}
\affiliation{Department of Physics and Astronomy, Bishop's 
University, 
2600 College Street, Sherbrooke, Qu\'ebec, Canada J1M~1Z7
}


%\date{\today}

\begin{abstract}

The uniqueness of the Kottler/Schwarzschild-de Sitter solution of the 
vacuum Einstein equations with positive cosmological constant is discussed 
and certain putative alternatives are shown to either solve different 
equations or to be the KSdS solution in disguise. A simultaneous no-hair 
and cosmic no-hair theorem for the KSdS geometry in the presence of an 
imperfect fluid is proved.

\end{abstract}

%\keywords{}

\maketitle

%\tableofcontents

\section{Introduction} 
\label{sec:1}
\setcounter{equation}{0}

The Jebsen-Birkhoff theorem \cite{Jebsen, Birkhoff} stating that the 
Schwarzschild geometry is the unique vacuum, spherically symmetric, and 
asymptotically flat solution of the Einstein equations is standard 
textbook material (see \cite{LivingReviews} for a review). Almost-Birkhoff 
theorems studying small deviations from 
spherical symmetry or vacuum have also been discussed 
\cite{almostBirkhoff, NziokiGoswamiDunsby2014}.  Relaxing the assumptions 
of the Jebsen-Birkhoff theorem to allow for an infinite distribution of 
matter leads to a variety of inhomogeneous universes \cite{Krasinskibook, 
Stephanietal}, which shows that there is no unique spherical solution with 
Friedmann-Lema\^itre-Robertson-Walker (FLRW) asymptotics. However, it is 
straightforward to extend the proof of the Jebsen-Birkhoff theorem to 
vacuum with a cosmological constant $\Lambda$ to deduce that the unique 
spherical solution of the vacuum Einstein equations in this case is the 
Kottler/Schwarzschild-de Sitter metric \cite{Kottler} (hereafter KSdS) if 
$\Lambda>0$ and the asymptotics are de Sitter, or the 
Schwarzschild-anti-de Sitter metric (SAdS) if $\Lambda<0$ and the 
asymptotics are anti-de Sitter. In locally static Schwarzschild-like 
coordinates $\left( T,R, \theta, \varphi \right)$ the KSdS metric has the 
form
\begin{eqnarray} 
ds^2 &=&-\left( 1-\frac{2m}{R}-H^2R^2 
\right) dT^2 +\frac{dR^2}{1-\frac{2m}{R}-H^2R^2} \nonumber\\
&&\nonumber\\
&\, &  +R^2 d\Omega_{(2)}^2 \,. \label{KSdS}
\end{eqnarray} 
Here $m$ and $H=\sqrt{\Lambda/3}$ are positive constants 
and $d\Omega_{(2)}^2=d\theta^2 +\sin^2 \theta \, 
d\varphi^2$ is the line element on the unit 2-sphere. The 
KSdS geometry plays the role of the prototypical black hole 
embedded in de Sitter space. The latter is extremely 
important for early universe inflation \cite{Lindebook, 
LiddleLyth} and is the late-time attractor of many dark 
energy and modified gravity models attempting to explain 
the current acceleration of the cosmic expansion 
\cite{AmendolaTsujikawabook} discovered in 1998 with type 
Ia supernovae. Likewise, anti-de Sitter space plays a 
prominent role in string theories and in the AdS/CFT 
correspondence \cite{Maldacena} which have been the subject of a large 
literature (see \cite{AdSCFTreviews} for recent reviews). It is 
surprising, 
therefore, that modern 
relativity textbooks do not mention the Jebsen-Birkhoff 
theorem in the presence of a cosmological constant, 
although occasionally one finds in the literature an explicit statement 
about 
the uniqueness of the Schwarzschild-(anti-)de Sitter space 
({\em e.g.}, 
\cite{Schmidt, NziokiGoswamiDunsby2014, 
FabianLasenby2015}). 
A proof of the 
Jebsen-Birkhoff theorem extended to include a non-vanishing 
$\Lambda$ is available in Synge's 1960  
textbook\footnote{Synge does not mention Kottler's   
paper \cite{Kottler}, nor does he refer to the KSdS solution as 
Schwarzschild-de Sitter but states that the metric is 
properly called ``Schwarzschild solution'' only when 
$\Lambda=0$.} on general relativity \cite{SyngeGR}. More 
mathematically sophisticated proofs of the uniqueness of 
the KSdS and SAdS space are contained in old and recent 
references \cite{recentproofs}.  Similar to the situation 
of the Schwarzschild solution, uniqueness implies that the 
KSdS and SAdS solutions are stable with respect to 
perturbations, the stability being established in 
Refs.~\cite{GuvenNunez90, BalbinotPoisson90, MellorMoss90, 
OtsukiFutamase91}. In spite of all this evidence, various 
works purport the existence of spherical solutions of the 
vacuum Einstein equations with $\Lambda>0$ which are 
alternatives to the KSdS one. This clearly cannot be true, 
or else these solutions must reduce to KSdS in 
disguise. There are also more general solutions of the 
Einstein equations representing central inhomogeneities 
embedded in FLRW  
spaces, which seem to reduce to alternatives to the KSdS 
solution in the special case when the FLRW ``background'' 
reduces to de Sitter. Again, this cannot be the case. 
Although these other authors presenting these solutions do 
not 
claim that they are alternatives to KSdS, nevertheless 
a situation was created which is unclear about the unique 
status of KSdS. To make things worse, enter alternative 
gravity. There is much interest in theories of gravity 
alternative to general relativity and in the study of their 
spherical solutions for various reasons. Although 
the Jebsen-Birkhoff theorem breaks down already in simple 
scalar-tensor gravity, some no-hair theorems persist and 
their relation with a positive cosmological constant has 
been discussed in the literature \cite{Hawking, nohair, 
SotiriouFaraoniPRL, RomanoPRL}. In particular, there are 
claims that spherical polytropic stars cannot match the 
KSdS exterior in scalar-tensor and $f(R)$ gravity 
\cite{Kaisa}, although the situation is still unclear 
in this regard \cite{SalvMaria}. Perhaps this happens because the KSdS 
solution is not adequate to describe inhomogeneous 
universes in these theories, but then one does not know 
which solution of the relevant field equations should be 
matched with the interior of a polytropic star, or with any 
local spherical object. It does not help these 
investigations if the situation is already confused in 
general relativity.  Our purpose here is to make clarity 
about the status of KSdS space in general relativity and to 
reveal putative alternatives as KSdS in disguise due to the 
use of non-standard coordinate systems, or to identify them 
as genuinely different solutions which obey different field 
equations with matter sources. We then present a new 
no-hair/cosmic no-hair theorem related to 
KSdS space in the presence of an imperfect fluid.  We use 
units in which Newton's constant $G$ and the speed of light 
$c$ are unity and we follow the notation of 
Ref.~\cite{Waldbook}.

\section{Uniqueness of the KSdS metric} 
\label{sec:2} 
\setcounter{equation}{0}

The most general spherically symmetric line element in four 
spacetime dimensions can be written in the form
\be
ds^2=-A^2(t,R)dt^2+B^2(t,R) dR^2 +R^2 d\Omega_{(2)}^2 \,.
\label{lineelement}
\ee
The vacuum Einstein equations
\be
G_{ab}=-\Lambda g_{ab}
\ee
then yield
\begin{eqnarray}
&& \frac{2\dot{B} }{RB}=0 \,, \label{JB1}\\
&&\nonumber\\
&& \frac{2B'}{B^3 R}-\frac{1}{B^2 R^2} +\frac{1}{R^2} = \Lambda  
\,, \label{JB2}\\
&&\nonumber\\
&& \frac{2A'}{A R}-\frac{B^2}{R^2} +\frac{1}{R^2} =-\Lambda B^2 
\,, \label{JB3}\\
&&\nonumber\\
&& \frac{A'B}{A}- B' - \frac{R B^2 
\ddot{B}}{A^2} +\frac{R \dot{A}\dot{B} B^2}{A^3} 
-\frac{RA' B'}{A} +\frac{ RA'' B}{A} \nonumber\\
&&\nonumber\\
&& =-\Lambda R B^3\,, \label{JB4}
\end{eqnarray}
where an overdot and a prime denote differentiation with 
respect to $t$ and $R$, respectively (the $(3,3)$ Einstein 
equation gives the same information as the $(2,2)$ 
equation). Using 
the consequence of Eq.~(\ref{JB1}) that $B=B(R)$, we drop 
the terms containing $\dot{B}$ or $\ddot{B}$ from 
Eq.~(\ref{JB4}). Equation~(\ref{JB2}) gives
\be
\left( \frac{R}{B^2} \right)'=  1 -\Lambda R^2 \,,
\ee
which is integrated to 
\be
B^2(R) = \frac{1}{1+\frac{C}{R}-\frac{\Lambda R^2}{3} } \,,
\ee
where $C$ is an integration constant. By imposing that one recovers 
the Schwarzschild solution for a mass $m$ as $\Lambda \rightarrow 
0$, one obtains $C=-2m$ and
\be
B^2= \frac{1}{1-\frac{2m}{R}- \frac{\Lambda R^2}{3} } \,.
\ee
Equation~(\ref{JB3}) now gives 
\be
\frac{2A'}{A} +\frac{1}{R}+ \frac{\Lambda R^2 -1}{
R\left( 1-\frac{2m}{R}-\frac{\Lambda R^2}{3} \right)} =0 \,,
\ee
which can be written as 
\be
\left( \ln A^2 \right)'=\left[ \ln \left( 1-\frac{2m}{R}- 
\frac{ \Lambda R^2}{3}  \right) \right] ' \,,
\ee
and integrates to 
\be
A^2(R)= \mbox{e}^{D(t)} \left( 1-\frac{2m}{R} - \frac{\Lambda 
R^2}{3}  \right)
\ee
where $D(t)$ is an integration function of time. At this stage, one is 
not entitled to assume that $\dot{A}=0$. However, by 
rescaling the time coordinate according to
\be
dT=\mbox{e}^{D(t)/2 } dt \,,
\ee
the spherically symmetric line element necessarily takes the static  
form
\begin{eqnarray}
ds^2 &=& -\left( 1-\frac{2m}{R}- \frac{\Lambda R^2}{3}  \right)dT^2 
+\frac{dR^2}{ 1-\frac{2m}{R}- \frac{\Lambda R^2}{3} } 
\nonumber\\
&&\nonumber\\
&\, & +R^2 d\Omega_{(2)}^2 \,.
\end{eqnarray}
This is the KSdS solution of the Einstein equations if $\Lambda 
>0$ (and then $H=\sqrt{\Lambda/3}$), the SAdS solution if $\Lambda 
<0$, and it reduces to the 
Schwarzschild solution if $\Lambda=0$.  The analysis of the 
spherical vacuum Einstein equations mirrors that, performed for 
$\Lambda=0$, which leads to the 
Jebsen-Birkhoff theorem in most relativity textbooks. If is 
therefore appropriate to speak of a generalized Jebsen-Birkhoff 
theorem when $\Lambda \neq 0$ and the KSdS solution is the unique 
solution of the vacuum Einstein equations with positive 
cosmological constant in spherical symmetry.

\section{Putative alternatives to KSdS}
\label{sec:3}
\setcounter{equation}{0}

Let us turn now to examining spherically symmetric 
solutions of the vacuum Einstein equations with $\Lambda>0$ 
which have been proposed as alternatives to the KSdS one, 
and to metrics which apparently contain alternatives to 
KSdS as special cases. Some ambiguity has been generated by 
the fact that these geometries have been presented in 
various coordinate systems, and different foliations of the 
KSdS spacetime can emphasize very different features ({\em 
e.g.}, \cite{Podolsky}).

\subsection{Abbassi-Meissner proposal}

Abbassi \cite{Abbassi} and, ten years later, Meissner 
\cite{Meissner} reported the following metric as a new 
alternative to the KSdS geometry (here we adopt the 
notation of \cite{Meissner}):
\be
ds^2 = -f(t,r)dt^2 +\frac{ \mbox{e}^{2Ht} }{f(t,r)} \, dr^2 + \,  
\mbox{e}^{2Ht} r^2 d\Omega_{(2)}^2 \,, \label{Abbassi}
\ee
where   
\be
f(t,r)= h(t,r)+\sqrt{ h^2(t,r)+H^2 r^2 \, \mbox{e}^{2Ht} } 
\,,
\ee
and
\be  
h(t,r) = \frac{1}{2} \left( 1-H^2 r^2  \, \mbox{e}^{2Ht} - 
\frac{2m}{r} \, \, \mbox{e}^{-Ht} \right) \,,
\ee
$m$ is a constant mass parameter and $H$ is the Hubble 
constant of the de Sitter background given by 
$H^2=\Lambda/3$. The areal radius of this spherically 
symmetric geometry is $ R(t,r)= a(t) \, r =\, 
\mbox{e}^{Ht}r$. Making use of the relation between 
differentials $ dr=a^{-1}\left( dR -HRdt\right)$, one 
rewrites the line element~(\ref{Abbassi}) in terms of the 
areal radius as
\begin{widetext}
\begin{eqnarray}
ds^2 &=& -2h(0,R) dt^2 -\frac{2HR}{f(0,R)}\, dtdR 
+\frac{dR^2}{f(0,R)} + R^2 d\Omega_{(2)}^2  \label{Abbassi2}\\
&&\nonumber\\
&=& - \left( 1-\frac{2m}{R}-H^2R^2 \right) dt^2 
\nonumber\\
&&\nonumber\\
&\, & -\frac{4HR}{   1-\frac{2m}{R}-H^2R^2+\sqrt{
\left(1-\frac{2m}{R}-H^2R^2 \right)^2+4H^2R^2}  }\, dtdR 
\nonumber\\
&&\nonumber\\
&\, & + 
\frac{2dR^2}{ 1-\frac{2m}{R}-H^2R^2 +\sqrt{
\left(1-\frac{2m}{R}-H^2R^2 \right)^2+4H^2R^2} } 
+ R^2 d\Omega_{(2)}^2 \,. 
\end{eqnarray}
\end{widetext}
By introducing a new time coordinate $T$ defined by 
\be
dT=dt+\beta(t, R) dR \,,
\ee
with $\beta (t,R) $ a function to be determined, and 
\be
A_0(R)\equiv 1-\frac{2m}{R}-H^2R^2 =2h(0, R)\,,
\ee
one obtains
\begin{eqnarray}
ds^2 &=& -A_0 dT^2 +\left( -A_0\beta^2 
+\frac{4HR\beta}{A_0+\sqrt{A_0^2+4H^2R^2}} \right. 
\nonumber\\
&&\nonumber\\
&\, & \left. +\frac{2}{ 
A_0+\sqrt{A_0^2+4H^2R^2}} \right) dR^2 
+R^2 d\Omega_{(2)}^2 \nonumber\\
&&\nonumber\\
&\, & + 2\left( \beta A_0-\frac{2HR}{ 
A_0+\sqrt{A_0^2+4H^2R^2}} \right) 
dTdR \,.\nonumber\\
&&
\end{eqnarray}
By setting
\be
\beta(R)= \frac{2HR}{ A_0\left( A_0+\sqrt{A_0^2+4H^2R^2} 
 \, \right)} 
\ee
the cross-term in $dTdR$ is eliminated and the line element  
assumes the diagonal and locally static form
\begin{eqnarray}
ds^2 &=& -A_0(R)dT^2 +\frac{2}{ A_0(R) 
+\sqrt{A_0^2(R)+4H^2R^2}} 
\nonumber\\
&&\nonumber\\
&\, & \cdot \left[ 
1+\frac{2H^2 R^2}{ A_0 \left( 
A_0+\sqrt{A_0^2+4H^2R^2}\right) } 
\right] dR^2 +R^2 d\Omega_{(2)}^2 \,,\nonumber\\
&&
\end{eqnarray}
which is not of the KSdS form. {\em A posteriori} one can 
check that 
$dT=dt+\beta dR$ is an exact differential ({\em i.e.}, the time 
coordinate $T$ is well defined) 
by noting that it is closed, 
\be
\frac{\partial (1)}{\partial R}= 0= 
\frac{\partial \beta}{\partial t} \,.
\ee 
Although claiming a new solution alternative to the KSdS 
one, Abbassi \cite{Abbassi} mentions a  
coordinate transformation that brings the line 
element~(\ref{Abbassi}) to the standard KSdS 
form,  but this coordinate change fails to do so. Moreover, this author 
ascribes different physical meanings to the same geometry 
described in different 
coordinate systems. The geometry, however, must be 
coordinate independent. In particular, the static 
character of the metric is shown by the existence of a 
timelike Killing vector field. In spite of what stated 
in~\cite{Abbassi, Meissner} the diagonal metric~(\ref{Abbassi}) 
does not solve the vacuum Einstein 
equations $R_{ab}=\Lambda g_{ab}$ but it is generated by 
matter sources. For example, there is a radial mass flow 
given by
\begin{widetext}
\begin{eqnarray}
T_{01}&=&\frac{1}{8\pi} \, 
\frac{d\left( \ln B^2 \right)}{dT} 
= \frac{H\dot{H}R^2}{ 2\pi \left(A_0 
+\sqrt{ A_0^2+4H^2R^2}\right) \sqrt{ A_0^2+4H^2R^2} } 
\nonumber\\
&&\nonumber\\
&\, & \cdot \left\{ -1 + \frac{ A_0\left(  A_0  
\sqrt{ A_0^2+4H^2R^2} +A_0^2 +2H^2 R^2\right) 
}{
\sqrt{ A_0^2+4H^2R^2} \left[  A_0\left(A_0
+\sqrt{ A_0^2+4H^2R^2}\right) +2H^2R^2 \right] } \right\} 
\,.
\end{eqnarray}
\end{widetext}

\subsection{McVittie and generalized McVittie solutions}

The McVittie solution was originally introduced to model the 
effect of the cosmological expansion on local systems 
\cite{McVittie} and has 
been the subject of much recent literature 
\cite{FaraoniJacques, McVittielit, Roshina, mylastbook}. It 
represents a 
central inhomogeneity (possibly a black hole)  embedded in a  
FLRW space. The source for the 
exterior McVittie metric is a fluid with energy density 
$\rho(t)$ 
which depends only on time, and pressure $P(t,r)$ which depends on 
both time and radius.  The line element can be 
cast in the form \cite{Arakida, Roshina}
\begin{eqnarray}
ds^2 &=& -\left[ 1-\frac{2m}{R}-H^2(t)R^2 \right] dt^2 
-\frac{2H(t) R}{\sqrt{1-2m/R}} \, dtdR \nonumber\\
&&\nonumber\\
&\, & +\frac{dR^2}{1-2m/R}  +R^2d\Omega_{(2)}^2 
\,,\label{Arakida}\\ 
&&\nonumber
\end{eqnarray}
where $m$ is a positive constant related to the mass of the central 
object and $H(t)$ is the Hubble parameter of the FLRW space in 
which this object is embedded.  When the FLRW 
``background'' reduces to de Sitter, $H=$~const., the 
transformation to the 
coordinate $T$ given by 
\be
dT=dt+ \frac{HR \, dR}{ \sqrt{1-\frac{2m}{R}} \left( 
1-\frac{2m}{R}-H^2R^2 
\right) }
\ee
reduces the metric to the KSdS form~(\ref{KSdS}). 
Therefore, the McVittie metric with $H=$~const. is not an alternative to 
KSdS but it contains it as a special case. 

In the literature there is also a class of ``generalized McVittie 
solutions'' in which, contrary to the original McVittie one, 
there is a spacelike radial heat flow $q^{\mu}=\left( 
0,q,0,0 \right)$ \cite{FaraoniJacques, 
Gaoetal}. McVittie spaces are also solutions of cuscuton theory (a 
special case of Ho\v{r}ava-Lifschitz gravity \cite{Afshordi}) and 
generalized McVittie 
spaces are also solutions of Horndeski gravity and shape dynamics 
\cite{Horndeski}. They are substantially more complicated than the 
McVittie one, but they also reduce to the KSdS geometry when the 
background is de Sitter \cite{FaraoniJacques, Gaoetal}, in which 
case the spacelike radial energy flow $q^a$ vanishes.

\subsection{Non-rotating Thakurta solution}

The Thakurta solution of the Einstein equations \cite{Thakurta} 
describes a rotating black hole embedded in a FLRW universe. 
When the angular momentum 
is set to zero and the cosmological background is chosen to be de 
Sitter, one obtains an apparent alternative to KSdS but 
this is not the case, as explained below. The non-rotating Thakurta 
solution was 
recently analyzed in detail  in \cite{MelloMacielZanchin16}, see 
also \cite{Culetu, Page}. The line element is 
\begin{eqnarray}
ds^2 &=& a^2(\eta) \left[ -\left( 1-\frac{2m}{r} \right) d\eta^2 
+ \frac{dr^2}{  1-2m/r } + r^2 d\Omega_{(2)}^2 \right] 
\label{Thakurta1}\nonumber\\
&&\nonumber\\
&=&  -\left( 1-\frac{2m}{r} \right) dt^2 
+ \frac{a^2 dr^2}{  1-2m/r } + a^2r^2 d\Omega_{(2)}^2 
\,,\label{Thakurta2}
\end{eqnarray}
where $a(\eta)$ is the scale factor of the FLRW background, 
$\eta$ and $t$ are its conformal and comoving times, respectively, 
with $dt=ad\eta$, and $m$ is a constant mass parameter. 
The line element~(\ref{Thakurta1}) is manifestly conformal to the 
Schwarzschild one. 
By using the areal radius $R(t,r)=a(t)r$ and the relation 
between differentials $dr=\frac{dR}{a} - HR d\eta$ (where $H\equiv 
\dot{a}/a$ and an overdot denotes differentiation with respect to 
the comoving time $t$), the line element is rewritten as 
\begin{eqnarray}
ds^2 &=& -\left( 1-\frac{2M(t)}{R} -\frac{H^2R^2}{ 1-2M(t)/R }   
\right) dt^2 \nonumber\\
&&\nonumber\\
&\, & + \frac{dR^2}{  1- 2M(t)/R} 
-\frac{2HR}{  1-2M(t)/R }\, dtdR + R^2 d\Omega_{(2)}^2 
\,,\nonumber\\
&& \label{Thakurtax}
\end{eqnarray}
where 
\be
M (t) \equiv ma(t) \,. 
\ee
The cross-term in $dtdR$ can be eliminated from this line element 
\cite{mySultanaDyer}. We use  $A (t, R) 
\equiv 1-2M/R=1-2m/r$ and a new time coordinate $T$ defined by
\be
dT= \frac{1}{F} \left( dt +\frac{HR}{ A^2-H^2R^2 }\, dR \right)
\ee
where $F(t, R)$ is an integrating factor satisfying
\be
\frac{\partial}{\partial R}\left( \frac{1}{F} \right)= 
\frac{\partial}{\partial t} \left( \frac{HR}{F\left( A^2-H^2R^2 
\right)} \right) \label{Thakurtaintegratingfactor}
\ee
to guarantee that $dT$ is an exact differential.  Straightforward 
manipulations bring the line 
element to the diagonal gauge
\begin{eqnarray}
ds^2 &= & -\left( 1-\frac{2M}{R}- \frac{H^2R^2}{1-\frac{2M}{R}} 
\right) F^2 dT^2 \nonumber\\
&&\nonumber\\   
&\, & +\frac{dR^2}{1-\frac{2M}{R} 
-\frac{H^2R^2}{1- 2M/R } } 
  +  R^2 d\Omega_{(2)}^2 \,.\label{ThakurtaDiag}
\end{eqnarray}
Using the form~(\ref{Thakurta2}) of the metric, the Einstein 
equations give \cite{MelloMacielZanchin16} 
\begin{eqnarray}
{G_0}^0= 8\pi {T_0}^0 &=& -\frac{3H^2}{A} \,,\\
&&\nonumber\\
{G_1}^0 = 8\pi {T_1}^0 &=& - \frac{2mH}{r^2 A^2 } 
\,,\label{efe10}\\ 
&& \nonumber\\
{G_1}^1= 8\pi {T_1}^1 &=& 8\pi {T_2}^2 = 8\pi {T_3}^3 = 
-\frac{1}{A} \left(H^2 + \frac{2\ddot{a}}{a} \right) \,.\nonumber\\
&& 
\end{eqnarray}
Assume a de Sitter background with $H=\sqrt{\Lambda/3}$ and 
$a(t)=a_0 \, \mbox{e}^{Ht}$; then the  time-radius 
Einstein equation~(\ref{efe10}) satisfied by the 
non-rotating Thakurta 
solution clearly cannot reduce to the 
corresponding equation satisfied by the KSdS metric, which would 
instead give $ 8\pi {T_1}^0 =-\Lambda {g_1}^0 =0$ (the 
vanishing of ${T_1}^0$ means that, because the 
cosmological constant is repulsive, it does not accrete 
onto a black hole and there is no radial energy flow). The 
two 
equations only coincide in the trivial cases when $m=0$  (de 
Sitter space) or when $a=$~const. (Minkowski background). These 
two equations 
cannot coincide because, as stated clearly in 
\cite{MelloMacielZanchin16, Culetu}, the 
source of the non-rotating Thakurta geometry is not a perfect 
fluid, to which 
the cosmological constant can be reduced, but is instead an 
imperfect one with a spacelike radial heat flow which has  
components $q_{\mu}=\left( 0, 
- 2m\dot{a} a A^{-3/2}/r^2 , 0, 0 \right)$ in 
coordinates $\left( t, r, \theta, \varphi 
\right)$ \cite{MelloMacielZanchin16}. 

It has been shown in Refs.~\cite{fatePLB, mylastbook} that the 
non-rotating Thakurta solution is the late time  attractor of 
generalized McVittie solutions, but these 
references\footnote{Ref.~\cite{Culetu}  studied the same 
geometry for different purposes and did not identify it with the 
Thakurta solution.} did not recognize the geometry as   
a special case of the less known Thakurta solution and  
called it ``comoving 
mass solution'' instead. The non-rotating Thakurta solution is also 
the limit to general 
relativity of a class of solutions of 
Brans-Dicke theory found in Ref.~\cite{CMB} as the Brans-Dicke 
parameter $\omega \rightarrow \infty$ \cite{fatePLB, mylastbook}.

\subsection{Castelo Ferreira metric}

Another line element which resembles, or even 
reduces to some of the previous ones for special 
parameter values was introduced by 
Castelo Ferreira \cite{Castelo} 
\begin{eqnarray}
ds^2 &=& -\left[ 1-\frac{2m}{R} -H^2R^2 \left( 
1-\frac{2m}{R} \right)^{\alpha} \right] dt^2 
+\frac{dR^2}{1-\frac{2m}{R}} \nonumber\\
&&\nonumber\\
&\, & -2HR \left( 1-\frac{2m}{R} \right)^{ \frac{ 
\alpha-1}{2}} dtdR +R^2 d\Omega_{(2)}^2 \,, \label{Castelo}
\end{eqnarray}
where $\alpha$ and $ m$ are constants and 
$H=H(t)$ is the 
Hubble parameter of the FLRW ``background''.  
This geometry does not satisfy the vacuum 
Einstein equations $G_{ab}=-\Lambda g_{ab}$ but is sourced 
by an imperfect fluid which has different tangential and 
radial pressures if $\alpha \neq 0$ \cite{Castelo}. The 
metric~(\ref{Castelo})  
reduces to 
the McVittie metric in the form ~(\ref{Arakida}) when 
$\alpha=0$ (in which case the two pressures coincide). In 
spite of superficial similarities, it does 
not reduce to the non-rotating Thakurta 
solution~(\ref{Thakurtax}) for $\alpha=-1$. 
Similarities and differences may be misleading 
because they depend on the coordinates adopted. Let us 
change the time coordinate $t\rightarrow T$, where $T$ is 
defined by
\be
dT=\frac{1}{F} \left( dt+\beta dR \right) \,,
\ee
where $1/F$ is an integrating factor and $\beta(t, R)$ is a 
function to be determined. The line element~(\ref{Castelo}) 
becomes
\begin{widetext}
\begin{eqnarray}
ds^2 &=& -\left[ 1-\frac{2m}{R} -H^2R^2 \left( 
1-\frac{2m}{R} \right)^{\alpha} \right] F^2 dT^2 
\nonumber\\
&&\nonumber\\
&\, & + \left\{ -\left[ 1-\frac{2m}{R}-H^2R^2 \left( 
1-\frac{2m}{R} \right)^{\alpha} \right]\beta^2 +\frac{1}{ 
1-\frac{2m}{R} } +2HR\beta \left( 1-\frac{2m}{R} \right)^{ 
\frac{\alpha-1}{2}} \right\}dR^2 \nonumber\\
&&\nonumber\\
&\, & +2F \left\{  
\left[ 1-\frac{2m}{R}-H^2R^2 \left(
1-\frac{2m}{R} \right)^{\alpha} \right]\beta
-HR \left(1-\frac{2m}{R} \right)^{\frac{
\alpha-1}{2}} \right\}  dTdR +R^2 d\Omega_{(2)}^2 \,. 
\label{Castelo1}
\end{eqnarray}
\end{widetext}
By setting 
\be 
\beta (t,R) = \frac{ HR \left(1-\frac{2m}{R} 
\right)^{\frac{
\alpha-1}{2}} }{1-\frac{2m}{R}-H^2R^2 \left(
1-\frac{2m}{R} \right)^{\alpha} }
\ee
the cross-term in $dTdR$ is eliminated and one obtains the 
line element in the diagonal gauge
\begin{eqnarray}
ds^2 &=& -\left[ 1-\frac{2m}{R} -H^2R^2 \left( 
1-\frac{2m}{R} \right)^{\alpha} \right] F^2 dT^2 
\nonumber\\
&&\nonumber\\
&\, & + \frac{dR^2}{ 
1-\frac{2m}{R}-H^2R^2 \left( 1-\frac{2m}{R} 
\right)^{\alpha} } +R^2 d\Omega_{(2)}^2 \,. \nonumber\\
&& \label{CasteloDiag}
\end{eqnarray}
If the background is de Sitter then $H=$~const., 
$\beta=\beta(R)$, and $F=1$, and the line 
element~(\ref{CasteloDiag}) reduces to the non-rotating 
Thakurta solution~(\ref{ThakurtaDiag}) for $\alpha=-1$ and 
to the KSdS form~(\ref{KSdS}) (which is a special case 
of McVittie) for $\alpha=0$.  
It is clear, however, that in the general case the geometry 
is different from the KSdS one.

\section{Simultaneous baldness and cosmic baldness}
\label{sec:4}
\setcounter{equation}{0}

Cosmic no hair theorems state that, with a few exceptions 
(Bianchi models which are overdense and collapse before the 
cosmological constant can come to dominate the dynamics), 
de Sitter space is an attractor in the late time dynamics 
of the universe \cite{Waldtheorem}. Similarly, under 
reasonable conditions, no-hair 
theorems for black holes exclude the possibility of fields 
in the exterior spacetime of black holes which would make 
the geometry deviate from Schwarzschild \cite{nohair}. Since the KSdS 
geometry brings together black hole physics and de Sitter cosmology, 
presumably simultaneous no-hair and cosmic 
no-hair results, pointing to the KSdS spacetime as the final 
attractor state, should be valid in the presence of a 
positive cosmological constant, spherical symmetry, and a 
central inhomogeneity. This idea is supported by the 
uniqueness of the KSdS solution {\em in vacuo} and  by its 
perturbative stability \cite{GuvenNunez90, 
BalbinotPoisson90, MellorMoss90, OtsukiFutamase91}. In the 
following we derive a non-perturbative result in this 
direction which is motivated by the presence of imperfect 
fluids in the solutions of the Einstein equations discussed 
in the previous sections.

Consider the Einstein equations with matter
\be
G_{ab}=-\Lambda g_{ab} +8\pi  T_{ab} \label{efe}
\ee
and assume spherical symmetry, in which case the line 
element is given by Eq.~(\ref{lineelement}). Assume that 
the solution of the Einstein equations is asymptotically de 
Sitter, that is, that there is a de Sitter-like 
cosmological horizon of areal radius $R_H$ and the 
solution of the Einstein equations~(\ref{efe}) reduces 
to~(\ref{KSdS}) as\footnote{The coordinates $\left( t, R, 
\theta, \varphi \right)$ are expected to break down when 
$R>R_H$ or when $R$ becomes smaller than the black hole 
horizon $R_{BH}$ that may be present. Outside of the region 
$R_{BH}\leq R \leq R_H$, the 
geometry is not expected to be locally static, as in KSdS 
space.} $R \rightarrow R_H^{-}$. The Einstein equations are 
now
\begin{eqnarray}
&& \frac{\dot{B} }{BR}=4\pi  T_{01} \,, \label{SS1}\\
&&\nonumber\\
&& A^2 \left( \frac{2B'}{B^3 R}-\frac{1}{B^2 R^2} +\frac{1}{R^2} 
\right) = \Lambda A^2 + 8\pi  T_{00} \,, \label{SS2}\\
&&\nonumber\\
&& \frac{2A'}{A R}-\frac{B^2}{R^2} +\frac{1}{R^2} =-\Lambda B^2 
+8\pi  T_{11} \,, \label{SS3}\\
&&\nonumber\\\
&&  \frac{A'B}{A}- B' - \frac{R B^2 
\ddot{B}}{A^2} +\frac{R \dot{A}\dot{B} B^2}{A^3} 
-\frac{RA' B'}{A} +\frac{ RA'' B}{A} \nonumber\\
&&\nonumber\\\
&& = \left( -\Lambda R^2 +8\pi  T_{22} \right) \frac{B^3}{R}  \,. 
\label{SS4}
\end{eqnarray}
Further assume that matter is described by an imperfect 
fluid with constant equation of state and a purely spatial 
radial heat flow (of the kind considered in the previous 
section),
\be
T_{ab}=\left( P+\rho \right) u_a u_b +P g_{ab} +q_a u_b 
+q_b u_a \,,\label{imperfect}
\ee
\be
P=w\rho \,, \;\;\;\;\; w=\mbox{const.} \,,
\ee
\be
u^a u_a=-1 \,, \;\;\;\;\;\; q^c u_c=0  \,.
\ee
The fluid 4-velocity and the radial energy flow have 
components
\begin{eqnarray}
u^{\mu} &=&\left( |A|^{-1}, 0,0,0 \right) \,, \;\;\;\;
u_{\mu}=\left( -|A|, 0,0,0 \right) \,, \\
&&\nonumber\\
q^{\mu}&=&\left( 0, q, 0,0 \right) \,, \;\;\;\;\;\;
q_{\mu}=\left( 0, B^2 q, 0,0 \right) \,.
\end{eqnarray}
The components of the stress-energy 
tensor~(\ref{imperfect}) are 
\begin{eqnarray}
T_{00} &= & A^2 \rho \,,\label{TTT00}\\
&&\nonumber\\
T_{01} &= & -|A|B^2 q \,,\label{TTT01}\\
&&\nonumber\\
T_{11} &= & B^2 P \,,\label{TTT11}\\
&&\nonumber\\
T_{22} &= & R^2 P \,,\label{TTT22}\\
&&\nonumber\\
T_{33} &= &  R^2 P \sin^2 \theta  \,. \label{TTT33}
\end{eqnarray}
$T_{01}>0$ and $q<0$ correspond to radial inflow, while 
$T_{01}<0$ and $q>0$ to outflow. 

In the case of inflow $q<0$, Eq.~(\ref{SS1}) yields
\be
(B^2)^{\centerdot} =-8\pi  |A|B^4 R q >0 \,, \label{Bdot}
\ee
therefore the metric component $B^2=g_{11}$ increases with 
time. Assuming the metric coefficients to be continuous 
and differentiable, there are then two possibilities: 
either $B^2(t,R) \rightarrow +\infty$ 
for any fixed $R$ as $t\rightarrow +\infty$ (or as $t\rightarrow 
t_{max}$ if there is a 
singularity at a finite future $t_{max}$), 
or $B^2(t, R)$ has  an horizontal asymptote as $t\rightarrow +\infty$. 

Let us consider the first case. The apparent horizons are 
located by the covariant equation $\nabla^c R \nabla_c 
R=0$, equivalent to $1/B^2=0$ in the coordinates used. If 
$B^2 \rightarrow +\infty $ as $t \rightarrow +\infty$ or as $t\rightarrow 
t_{max}$, then 
at late times all points of space at any value of $R$ lie 
arbitrarily close to an apparent horizon. This situation is 
familiar in cosmology: it corresponds to a phantom universe in 
which there is a Big Rip singularity at a finite time $t_{max}$ and the 
apparent horizon (which has areal radius 
$R_{AH}=H^{-1}$ in a spatially flat FLRW cosmos \cite{mylastbook}) shrinks 
around a comoving  
observer because the expansion of the universe super-accelerates, {\em 
i.e.}, $\dot{H}=-4\pi \left(P+\rho \right)>0$ \cite{Caldwell}. By 
contrast, in a de Sitter space the Hubble parameter $H$ remains constant 
although the expansion itself accelerates, $\ddot{a}>0$. In a universe 
dominated by non-phantom dark energy (other than the cosmological 
constant), it is instead $\dot{H}<0$ while $\ddot{a}>0$. These phantom 
asymptotics contradict our assumption of de Sitter asymptotics and, 
therefore, we discard this possibility.

There remains the case in which $B^2(t, R)$ asymptotes to a function 
$B_0^2(R)$ of $R$   
as $t \rightarrow +\infty$. In this case $\dot{B} 
\rightarrow 0$ as $t \rightarrow +\infty$ (which also 
implies that the apparent horizons located by the equation 
$1/B^2=0$ become less and less dynamical). Then 
Eq.~(\ref{Bdot}) implies that the radial flow $q 
\rightarrow 0$ as $t \rightarrow +\infty$. In conjunction 
with Eq.~(\ref{TTT00}), the differentiation of 
Eq.~(\ref{SS2}) yields
\be 
8\pi \dot{\rho}= \frac{2}{R} \left( \frac{B'}{B^3} 
\right)^{\centerdot} 
-\frac{1}{R^2} \left( \frac{1}{B^2} \right)^{\centerdot} 
\rightarrow 
0  \;\;\;\;\;\mbox{as} \;\; t \rightarrow +\infty \,.
\ee
The assumption that $P=w\rho$ with constant $w$ (or with $w=w(R)$) then 
implies that also $\dot{P} \rightarrow 0$ as $ t 
\rightarrow +\infty$. 
Equations~(\ref{TTT11}) and~(\ref{SS3}) give 
\be
8\pi  \dot{P}= \frac{2}{R} \left( \frac{A'}{AB^2} 
\right)^{\centerdot} 
+\frac{1}{R^2} \left( \frac{1}{B^2} \right)^{\centerdot} \approx 
\frac{2}{RB^2} \left(   \frac{A'}{A} \right)^{\centerdot} 
\rightarrow 0
\ee
as $t \rightarrow +\infty $. Therefore, also $A^2 $ becomes 
time-independent, and the metric becomes static as  $t 
\rightarrow +\infty $.

To make progress, consider the covariant conservation equation 
$\nabla^b T_{ab}=0$ for the imperfect fluid stress-energy 
tensor~(\ref{imperfect}), which yields 
\begin{eqnarray}
&&u_au^b \nabla_b \left( P+\rho \right) 
+\left[ \left(P+\rho \right) u_a +q_a \right] \nabla^b u_b 
\nonumber\\
&&\nonumber\\
&&+ \left[ \left(P+\rho \right) u_b +q_b \right] \nabla^b u_a
+\nabla_a P +u^b \nabla_b q_a +u_a \nabla^b q_b=0 \,.\nonumber\\
\end{eqnarray}
Projecting this equation onto the time direction $u^a$ of 
comoving observers and using the orthogonality of 
4-velocity and 4-acceleration $u^a \nabla_b u_a=0$, one 
obtains 
\be
-\dot{\rho} -\left(P+\rho \right)  \nabla^b u_b +
u^a q^b \nabla_b u_a + u^a u^b \nabla_b q_a 
-\nabla^b q_b =0 \,.
\ee
At late times $q^c$ and $\dot{\rho}$ disappear from this equation, 
as we have deduced above, leaving
\be
\left(P+\rho \right)\nabla^b u_b \simeq 0 \,.
\ee
In general $\nabla^b u_b $ is different from zero (indeed, 
since the geometry must be asymptotically de Sitter at 
large radii, $\nabla^b u_b $ reduces to $3H>0$ there) and 
we are left with $P+\rho \rightarrow 0$ as $t \rightarrow 
+\infty $. Either the matter fluid reduces to a 
cosmological constant, in which case the vacuum uniqueness 
theorem for KSdS holds, or else both $\rho$ and $P=w\rho$ become 
subdominant and the cosmological constant dominates the 
expansion at late times while $\rho$ and $P$ become 
unimportant. Also in this case the solution reduces to 
KSdS.

If instead there is outflow $q>0$, then 
\be
(B^2)^{\centerdot} =-8\pi  |A|B^4 R q <0 
\ee
and, since $B^2 $ is bounded from below by zero and it 
decreases as $t\rightarrow +\infty$, it must have a 
horizontal asymptote with $B^2 (t, R) \rightarrow 
B_0^2(R)^{+}$ for any fixed $R$ as $t\rightarrow +\infty$. 
Then $\dot{B}\rightarrow 0$ and $q \rightarrow 0$. The 
reasoning made in the case with inflow is then repeated 
from this point, reaching the same conclusion. Hence it is 
proved that, assuming spherical symmetry, $\Lambda>0$ 
and spatial de Sitter asymptotics, and an imperfect fluid with 
constant equation of state and purely spatial radial energy 
flow, the late time solution of the Einstein equations must 
be the KSdS geometry.

As a special case, one can consider a perfect fluid by setting $q^a =0$. 
In this case $T_{01}=0$ and Eq.~(\ref{TTT01}) gives $B=B(R)$. It is then 
straightforward to prove that it must be $P=-\rho$ and that the KSdS 
geometry can be the only solution (this simple proof for a perfect 
fluid was already given in 
Ref.~\cite{FaraoniJacques}).

\smallskip
\section{Conclusions}
\label{sec:6}
\setcounter{equation}{0}

As seen in Sec.~\ref{sec:2}, the KSdS solution is the 
unique spherically symmetric solution of the vacuum 
Einstein equations with positive cosmological constant. 
This result is a simple generalization of the 
ordinary Jebsen-Birkhoff theorem \cite{Jebsen, Birkhoff}, 
which makes the same assumptions except that it assumes  
$\Lambda=0$, 
and goes hand-in-hand with the perturbative analyses which 
established the stability of the KSdS solution 
\cite{GuvenNunez90, BalbinotPoisson90, MellorMoss90, 
OtsukiFutamase91}. Putative alternative solutions of the 
Einstein equations under the same conditions either solve 
different equations (for example, including an imperfect 
fluid with spacelike radial flow) or are just the KSdS 
solution in disguise in an unusual coordinate system.

Since no-hair theorems reinforce the uniqueness of the 
Schwarzschild geometry \cite{nohair} and cosmic no-hair 
theorems establish that the de Sitter space is the unique 
late-time attractor in cosmology (with few exceptions 
\cite{Waldtheorem}), it is reasonable to expect that 
similar theorems should hold for the KSdS spacetime, which 
is usually interpreted as describing a Schwarzschild black 
hole embedded in de Sitter space. Such theorems would prove the 
uniqueness of the KSdS solution. We have proved a result of 
this kind by assuming spherical symmetry and the presence 
of an imperfect fluid with constant equation of state 
$P=w\rho$ and a purely spatial radial energy flow, which is 
a rather common ingredient in the construction of solutions 
of the Einstein equations representing spherical 
inhomogeneous universes ({\em e.g.}, \cite{FaraoniIsrael, 
FaraoniJacques, Gaoetal, MelloMacielZanchin16, 
McVittielit}). The theorem proved in Sec.~\ref{sec:4} does not contradict 
the previous statement of Sec.~\ref{sec:3} that the non-rotating Thakurta 
solution is the late-time attractor of generalized McVittie solutions 
\cite{fatePLB} because, in this case, the asymptotics are (time-dependent) 
FLRW and not de Sitter, which was one of the assumptions in our theorem. 
Simultaneous cosmic no-hair/no-hair theorems more general than the 
one in Sec.~\ref{sec:4} (possibly including anisotropy) will be 
investigated in the future.

\def\theequation{A.\arabic{equation}}\setcounter{equation}{0}

\begin{acknowledgments} 

We thank Jeremy Cot\'e and Thomas Gobeil for discussions. This work is 
supported by the Natural Sciences and Engineering Research Council of 
Canada. 

\end{acknowledgments}

%\bibliography{simplified}

\end{document}